\documentclass[aps, prd, nofootinbib, preprintnumbers, twocolumn, showpacs]{revtex4}
\usepackage{graphicx}
\usepackage{amsmath}
\usepackage{multirow}
\usepackage{bm}
\def\fsl#1{\setbox0=\hbox{$#1$}                 % set a box for #1
   \dimen0=\wd0                                 % and get its size
   \setbox1=\hbox{/} \dimen1=\wd1               % get size of /
   \ifdim\dimen0>\dimen1                        % #1 is bigger
      \rlap{\hbox to \dimen0{\hfil/\hfil}}      % so center / in box
      #1                                        % and print #1
   \else                                        % / is bigger
      \rlap{\hbox to \dimen1{\hfil$#1$\hfil}}   % so center #1
      /                                         % and print /
   \fi}                                         %

\newcommand{\Tr}{\mbox{Tr}}
\newcommand{\diag}{\mbox{diag}}
\newcommand{\VEV}[1]{\langle #1 \rangle}

\begin{document}
\title{Meissner screening masses in gluonic phase}
\author{Michio Hashimoto}
 \email{mhashimo@uwo.ca}
  \affiliation{
   Department of Applied Mathematics, 
   University of Western Ontario, 
   London, Ontario N6A 5B7, Canada}
\author{Junji Jia}
 \email{jjia5@uwo.ca}
   \affiliation{
    Department of Applied Mathematics, 
    University of Western Ontario, 
    London, Ontario N6A 5B7, Canada}
\date{\today}
\preprint{UWO-TH-07/16}
 
\begin{abstract}
A numerical analysis for the Meissner mass in the simplest gluonic phase
(the minimal cylindrical gluonic phase II) is performed
in the framework of the gauged Nambu-Jona-Lasinio model with 
cold two-flavor quark matter.
We derive Meissner mass formulae without using the numerical second 
derivative.
It is revealed that the gapless mode yields a characterized contribution
to the Meissner mass.
We also find that there are large and positive contributions from 
the tree gluon potential term to the transverse modes of gluons. 
It is shown that the simplest gluonic phase resolves
the chromomagnetic instability in a rather wide region.
\end{abstract}

\maketitle

\section{Introduction}

Quark matter at sufficiently high density and low temperature 
is expected to be in a color superconducting state driven by 
the BCS mechanism~\cite{Bailin:1983bm,Iwasaki:1994ij,CSC2}.
This is analogous to the electron Cooper paring
in a superconducting metal.
However, quarks, unlike electrons, have color and flavor
degrees of freedom as well as spin, so that 
the phase structure is quite rich.
In nature, deconfined quark matter might
exist in the interior of neutron stars~\cite{quark_star}.
Thus the dynamics of the color superconductivity 
has been intensively studied~\cite{review}. 

Bulk matter in compact stars should be in equilibrium under 
the weak interaction ($\beta$-equilibrium),
and be electrically and color neutral.
The electric and color neutrality conditions play 
a crucial role in the dynamics of the quark 
pairing~\cite{Iida:2000ha,Alford:2002kj,Steiner:2002gx,Huang:2002zd}. 
In addition, the strange quark mass cannot be neglected 
in moderately dense quark matter as in the compact stars.
Then a mismatch $\delta\mu$ between 
the Fermi momenta of the pairing quarks is induced.

As the mismatch $\delta\mu$ increases, the conventional color superconducting 
state tends to be destroyed. 
Before the complete destruction, however,
the Meissner mass of gluons turns to be imaginary
in the gapped (2SC) and gapless (g2SC) two-flavor color superconducting 
phases~\cite{Huang:2004bg}:
In the g2SC phase with the diquark gap $\Delta < \delta\mu$
the 8th gluon has an imaginary Meissner mass, while 
the Meissner masses for the 4-7th gluons are imaginary also in
the 2SC phase $\delta\mu < \Delta < \sqrt{2} \delta\mu$.
This chromomagnetic instability implies that there should exist
a more stable vacuum other than the 2SC/g2SC phase.
Later a chromomagnetic instability was found also 
in the three-flavor gapless color-flavor locked (gCFL) 
phase~\cite{Casalbuoni:2004tb,Alford:2005qw,Fukushima:2005cm}.
One of the central issues in this field is to establish
the genuine ground state for realistic values of $\delta\mu$.
Besides the gluonic phase~\cite{Gorbar:2005rx,Gorbar:2007vx}, 
a number of other candidates for the true vacuum have been proposed~\cite{Alford:2000ze,Bowers:2002xr,Reddy:2004my,Huang:2005pv,Hong:2005jv,Kryjevski:2005qq,Schafer:2005ym,Casalbuoni:2005zp,Rajagopal:2006ig,Mannarelli:2007bs,Gatto:2007ja}.

Connected with the chromomagnetic instability,
it was revealed that there appear tachyonic plasmons 
in the 4-7th and 8th gluonic channels~\cite{Gorbar:2006up}.
It clearly shows that the physical vectorial excitations
carry the instabilities and thus supports the scenario with 
gluon condensates (gluonic phase). 

It is also known that the physical diquark excitation
(the diquark Higgs mode) suffers from 
the Sarma instability~\cite{sarma} in the g2SC region,
which corresponds to the negative mass squared of 
the diquark Higgs at zero momentum.
Furthermore, it was found that the diquark Higgs mode
has a negative velocity squared~$v^2 < 0$ 
in the g2SC region~\cite{Hashimoto:2006mn}.
A similar instability is also discussed in 
Refs.~\cite{Iida:2006df,Giannakis:2006gg}.
This problem should be also resolved in the genuine ground state.

In Refs.~\cite{Gorbar:2005rx,Gorbar:2007vx},
the Ginzburg-Landau (GL) approach in the hard dense loop (HDL)
approximation was employed in the vicinity of 
$\delta\mu \approx \Delta/\sqrt{2}$.
Outside the scaling region around $\delta\mu \approx \Delta/\sqrt{2}$,
the self-consistent analysis by solving 
the gap equations and the neutrality conditions was recently performed
in Ref.~\cite{Hashimoto:2007ut}:
It is shown that the gluonic phase is actually realized 
in a wide region of the parameter space and it is energetically 
more favorable than the normal, 2SC/g2SC, and 
the single plane wave Larkin-Ovchinnikov-Fulde-Ferrell 
(LOFF)~\cite{Alford:2000ze,Giannakis:2004pf,Gorbar:2005tx} phases.
It is also found that the values of $\Delta$ and $\delta\mu$
in the gluonic phase are significantly different from 
those in the 2SC/g2SC phase.
It is noticeable that the values of the gluon condensate 
are large, say, ${\cal O}(\mbox{100--250MeV})$ in the almost whole
region where it exists.
On the other hand, the values of the color chemical potentials 
are relatively small.
For the earlier works in other approaches, see 
Refs.~\cite{Fukushima:2006su,Kiriyama:2006ui}.
The extension to the model with nonzero temperature is studied
in Ref.~\cite{Kiriyama:2007ng}.

In this paper, we examine whether or not the gluonic phase 
resolves the chromomagnetic instability.
We derive formulae for the Meissner mass without using
the numerical second derivative.
It turns out that the gapless mode gives a special contribution 
to the Meissner mass.
In the numerical analysis, we consider the gluonic phase with 
the simplest ansatz which is called the minimal cylindrical gluonic 
phase~II~\cite{Gorbar:2007vx,Hashimoto:2007ut}.
As a benchmark, we also analyze the single plane wave LOFF and
2SC/g2SC phases including the non-HDL corrections.

We find that in the gluonic phase the tree gluon potential term 
yields positive and large contributions to the Meissner masses of 
the transverse modes of gluons.
Actually, in the minimal cylindrical gluonic phase~II,
the chromomagnetic instability is resolved in the weak and 
intermediate coupling region, 
$65.4 \mbox{MeV} < \Delta_0  < 130 \mbox{MeV}$
for $\mu=400$MeV and $\Lambda=653.3$MeV, where 
$\mu$ and $\Lambda$ denote the quark chemical potential and 
the cutoff in the (gauged) Nambu-Jona-Lasinio (NJL) model, respectively. 
We here introduced the 2SC gap parameter $\Delta_0$ defined at
$\delta\mu=0$, which essentially corresponds to the four-diquark
coupling constant in the (gauged) NJL model.
In the intermediate and strong coupling region
$130 \mbox{MeV} < \Delta_0  < 160 \mbox{MeV}$, however,
the chromomagnetic instability occurs in the transverse modes of 
the 4th and 5th gluons.
Besides, in a small region around $\Delta_0 \simeq 150 \mbox{MeV}$,
the squared Meissner mass for the transverse mode of the 8th gluon 
becomes negative.
For the other modes, the chromomagnetic instability does not occur.
On the other hand,
the single plane wave LOFF phase resolves
the chromomagnetic instability only in the region 
$64.9 \mbox{MeV} < \Delta_0  < 80 \mbox{MeV}$.
This is consistent with the results within the HDL approximation
shown in Ref.~\cite{Gorbar:2005tx}.
Numerically, the 2SC phase suffers from the illness
in $134.6 \mbox{MeV} < \Delta_0 < 160 \mbox{MeV}$,
whereas the g2SC phase does in $\Delta_0 < 134.6 \mbox{MeV}$. 
We thus conclude that the situation is definitely improved 
even in the simplest gluonic phase.
If we consider more complicated gluonic phases such as 
the cylindrical gluonic phase~I~\cite{Gorbar:2005rx,Gorbar:2007vx} 
and/or the gluonic color-spin locked (GCSL) 
phase~\cite{Gorbar:2007vx,Hashimoto:2007ut},
the chromomagnetic instability might be completely removed.

The paper is organized as follows: 
In Sec.\ref{model}, the gauged NJL model is described. 
We also show the dynamical solutions of the gluonic, LOFF and 
2SC/g2SC phases.
In Sec.\ref{formulae-sub1}, we develop the formulae for 
the numerical calculation of the Meissner masses.
We numerically analyze the Meissner masses for the gluonic, LOFF and 
2SC/g2SC phases in Sec.\ref{formulae-sub2}.
Section~\ref{summary} presents the summary and discussions.

\section{Model}
\label{model}

We study the gauged Nambu-Jona-Lasinio (NJL) model with 
two light quarks.
We neglect the current quark masses and
the $(\bar{\psi}\psi)^2$-interaction channel.
The Lagrangian density is given by
\begin{eqnarray}
  {\cal L} &=& \bar{\psi}(i\fsl{D}+\bm{\mu}_0\gamma^0)\psi
  +G_\Delta \bigg[\,(\bar{\psi}^C i\varepsilon\epsilon^\alpha\gamma_5 \psi)
           (\bar{\psi} i\varepsilon\epsilon^\alpha\gamma_5 \psi^C)\,\bigg]
  \nonumber \\
&&  -\frac{1}{4}F_{\mu\nu}^{a} F^{a\,\mu\nu} ,
 \label{Lag}
\end{eqnarray}
with
\begin{equation}
  D_\mu \equiv \partial_\mu - i g A_\mu^{a} T^{a}, \quad
  F_{\mu\nu}^{a} 
  \equiv \partial_\mu A_\nu^{a} - \partial_\nu A_\mu^{a} +
  g f^{abc} A_\mu^{b} A_\nu^{c},
\end{equation}
where $\varepsilon$ and $\epsilon^{\alpha}$ are the totally 
antisymmetric tensors in the flavor and color spaces, respectively. 
We also introduced gluon fields $A_\mu^{a}$,
the QCD coupling constant $g$, 
the generators $T^{a}$ of $SU(3)$, 
and the structure constants $f^{abc}$.
The quark field $\psi$ is a flavor doublet and color triplet.
The charge-conjugate spinor is defined by
$\psi^C \equiv C \bar{\psi}^T$ with $C = i\gamma^2\gamma^0$.
We do not introduce the photon field.
On the other hand, the whole theory contains free electrons,
although we do not show them explicitly in Eq.~(\ref{Lag}).
In $\beta$-equilibrium, the chemical potential matrix $\bm{\mu}_0$ 
for up and down quarks is 
\begin{equation}
  \bm{\mu}_0 = \mu {\bf 1} - \mu_e Q_{\rm em}, 
 \label{mu}
\end{equation}
with ${\bf 1} \equiv {\bf 1}_c \otimes {\bf 1}_f$, and
$Q_{\rm em} \equiv {\bf 1}_c \otimes \diag(2/3,-1/3)_f$,
where
$\mu$ and $\mu_e$ are the quark and electron chemical potentials,
respectively.
(The baryon chemical potential $\mu_B$ is given by $\mu_B \equiv 3\mu$.)
The subscripts $c$ and $f$ mean that the corresponding matrices act on 
the color and flavor spaces, respectively.
Hereafter, we abbreviate the unit matrices, ${\bf 1}$, 
${\bf 1}_c$ and ${\bf 1}_f$, if it is self-evident. 
By introducing the diquark field
$\Phi^\alpha \sim i\bar{\psi}^C\varepsilon \epsilon^\alpha \gamma_5 \psi$,
we can rewrite the Lagrangian density (\ref{Lag}) as
\begin{eqnarray}
  {\cal L} &=& \bar{\psi}(i\fsl{D}+\bm{\mu}_0 \gamma^0)\psi
  - \frac{|\Phi^\alpha|^2}{4G_\Delta} 
  - \frac{1}{2}\Phi^\alpha
     [i\bar{\psi}\varepsilon\epsilon^\alpha \gamma_5 \psi^C]
 \nonumber \\ &&
  - \frac{1}{2}
    [i\bar{\psi}^C\varepsilon\epsilon^\alpha\gamma_5 \psi]\Phi^{*\alpha}
  - \frac{1}{4}F_{\mu\nu}^{a} F^{a\,\mu\nu} .
 \label{Lag_aux}
\end{eqnarray}

In the 2SC/g2SC phase,
we can choose the anti-blue direction without loss of generality, 
\begin{equation}
  \VEV{\Phi^r}=0, \quad  \VEV{\Phi^g}=0, \quad \VEV{\Phi^b}=\Delta ,
  \label{2SC}
\end{equation}
where the diquark condensate $\Delta$ is real.
In this basis, by imposing the color neutrality condition,
the color chemical potential $\mu_8$ is induced~\cite{Gerhold:2003js}.
We can interpret $\mu_8$ as the vacuum expectation value (VEV) of 
the time component of the 8th gluon.

Let us define the Nambu-Gor'kov spinor,
\begin{equation}
  \Psi \equiv \left(\begin{array}{@{}c@{}} \psi \\ \psi^C \end{array}\right) .
\end{equation}

The propagator inverse of $\Psi$ in the 2SC/g2SC phase is given by
\begin{equation}
  S^{-1}(P) = \left(
  \begin{array}{cc}
  [G_0^+]^{-1} & \Delta^- \\ \Delta^+ &  [G_0^-]^{-1} 
  \end{array}
  \right),
  \label{S-inv}
\end{equation}
with 
\begin{equation}
  [G_0^+]^{-1}(P) \equiv
  (p_0+\bar{\mu}-\delta\mu\tau_3-\mu_8{\bf 1}_b)\gamma^0
  -\vec \gamma \cdot \vec p, 
\end{equation}
\begin{equation}
  [G_0^-]^{-1}(P) \equiv
  (p_0-\bar{\mu}+\delta\mu\tau_3+\mu_8{\bf 1}_b)\gamma^0
  -\vec \gamma \cdot \vec p,
\end{equation}
and
\begin{equation}
  \Delta^- \equiv -i\varepsilon\epsilon^b\gamma_5\Delta, \;\;
  \Delta^+ \equiv \gamma^0 (\Delta^{-})^\dagger \gamma^0 =
  -i\varepsilon\epsilon^b\gamma_5\Delta,
\end{equation}
where $P^\mu \equiv (p_0,\vec p)$ is the energy-momentum four vector.
We also defined $\tau_3 \equiv \diag(1,-1)_f$,
${\bf 1}_b \equiv \diag (0,0,1)_c$, and
\begin{equation}
  \bar{\mu} \equiv \mu - \frac{\delta\mu}{3}+\frac{\mu_8}{3} , \qquad
  \delta \mu \equiv \frac{\mu_e}{2}.
  \label{def-mu} 
\end{equation}

The 2SC/g2SC phase is not the genuine ground state
in the region $\delta\mu > \Delta/\sqrt{2}$, 
because it suffers from the chromomagnetic instability.
A candidate to resolve the chromomagnetic instability is
the gluonic phase with gluon condensates~\cite{Gorbar:2005rx}.

Let us introduce the gluon condensates $\VEV{A_\mu^a} \ne 0$. 
When the space-component gluon condensates $\VEV{\vec A^a} \ne 0$
are incorporated into the theory,
the time-component VEVs of the gluon fields other than the 8th one
are generally induced as well.
We may interpret them as the color chemical potentials\footnote{
In an appropriate basis, the color chemical potentials can be reduced 
only into $\mu_3$ and $\mu_8$, 
because the color chemical potential matrix 
$\bm{\mu}_c \equiv \mu_a T^a$ ($a=1,2,\cdots,8$) 
is hermite and traceless.
In this case, however, the basis for the diquark field changes from
Eq.~(\ref{2SC}).}~\cite{Buballa:2005bv}:
\begin{equation}
  \mu_{\breve{a}} = g \VEV{A_0^{\breve{a}}},
  \; (\breve{a}=1,2,\cdots,7) , \;\;
  \mu_8= \frac{\sqrt{3}}{2} g\VEV{A_0^{8}} \, .
\end{equation}
The propagator inverse $S_g^{-1}$ of $\Psi$ including the gluon condensates
is written as
\begin{equation}
  S_g^{-1}(P) = \left(
  \begin{array}{cc}
  [G_{0,g}^+]^{-1} & \Delta^- \\ \Delta^+ &  [G_{0,g}^-]^{-1} 
  \end{array}
  \right) ,
  \label{Sg-inv}
\end{equation}
with
\begin{align}
&  [G_{0,g}^+]^{-1}(P) \equiv
  (p_0+\bm{\mu}_0)\gamma^0
  -\vec \gamma \cdot \vec p + g \VEV{\fsl{A}^a} T^a, \\[2mm]
&  [G_{0,g}^-]^{-1}(P) \equiv
  (p_0-\bm{\mu}_0)\gamma^0
  -\vec \gamma \cdot \vec p - g \VEV{\fsl{A}^a} (T^a)^T .
\end{align}

In the fermion one-loop approximation, 
the bare effective potential including both gluon and diquark condensates 
is given by
\begin{equation}
  V_{\rm eff}^{\rm bare} =
   \frac{\Delta^2}{4G_\Delta}
  +\frac{1}{4}F_{\mu\nu}^{a} F^{a\,\mu\nu}
  -\frac{\mu_e^4}{12\pi^2}
  -\frac{1}{2}\int\frac{d^4 P}{i(2\pi)^4}\Tr\ln S_g^{-1} ,
  \label{V_exp}
\end{equation}
where we added the free electron contribution.
Since the bare potential has a divergence, 
a counter term is required.
We take into account only differences of the free energies
with and without the chemical potentials.
We thus define the renormalized effective potential by
\begin{equation}
  V_{\rm eff}^R \equiv V_{\rm eff}^{\rm bare} - V_{\rm c.t.} ,
\end{equation}
with the counter term,
\begin{equation}
  V_{\rm c.t.} = -\frac{1}{2}\int\frac{d^4 P}{i(2\pi)^4}\Tr\ln 
  S_g^{-1} \Bigg|_{\mu=\mu_e=\mu_a=0,\Delta=0,\VEV{\vec A^a} \ne 0} \, . 
  \label{ct}
\end{equation}
In this prescription, even if we use the regularization scheme 
with the sharp cutoff $\Lambda$ for the loop integral, 
we can remove artificial mass terms of gluons 
like $\Lambda^2 \vec A_a^2$.

In general, we can reduce the 32 homogeneous gluon condensates
to 25 ones~\cite{Gorbar:2007vx}.
However the general case is quite complicated and hence it is
difficult to find the self-consistent solutions of the gap equations
and the color neutrality conditions for the 25 VEVs at present.

\begin{figure}[t]
 \begin{center}
 \resizebox{0.47\textwidth}{!}{\includegraphics{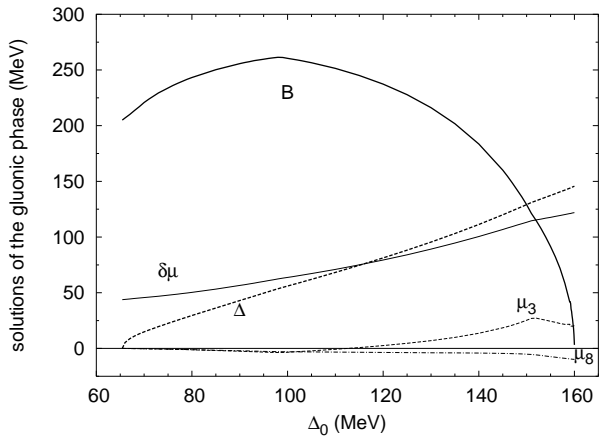}}
 \end{center}
 \caption{The dynamical solutions for 
          the minimal cylindrical gluonic phase II.
          The bold solid, bold dashed, thin solid, thin dashed and 
          thin dot-dashed curves represent the values of 
          $B$, $\Delta$, $\delta\mu$, $\mu_3$ and $\mu_8$, respectively.
          The values $\Lambda=653.3$ MeV and $\mu=400$ MeV were used.
          We also took $\alpha_s=1$.
          \label{sol_glu}}
\end{figure}

In this paper, we consider the minimal ansatz for the cylindrical gluonic 
phase II~\cite{Gorbar:2007vx,Hashimoto:2007ut},
\begin{equation}
  \mu_3 \equiv g\VEV{A_0^{3}}, \quad 
  \mu_8 \equiv \frac{\sqrt{3}}{2} g\VEV{A_0^{8}}, \quad
  B \equiv g \VEV{A_z^{6}} \, . 
\end{equation}
In order to make the physical meaning of the gluon condensates clearer,
it is convenient to use the unitary gauge in which all gauge dependent
degrees of freedom are removed.
We shall fix the gauge as follows~\cite{Gorbar:2007vx}:
\begin{equation}
  \Phi^r \equiv 0, \quad  \Phi^g \equiv 0, \quad 
  \mbox{Im}\Phi^b \equiv 0 , 
  \label{U-gauge1}
\end{equation}
and
\begin{equation}
  A_z^4 \equiv 0, \quad   A_z^5 \equiv 0, \quad   A_z^7 \equiv 0 \, . 
  \label{U-gauge2}
\end{equation}

As a benchmark, we also consider the single plane-wave 2SC-LOFF 
phase~\cite{Alford:2000ze,Giannakis:2004pf,Gorbar:2005tx},
\begin{equation}
  \VEV{\Phi^{r}} = \VEV{\Phi^{g}} = 0, \quad
  \VEV{\Phi^{b}} = \Delta e^{-2i\vec q \cdot \vec x} \, .
  \label{LOFF}
\end{equation}
Introducing the quark field $\psi'=e^{i\vec q \cdot \vec x}\psi$
with the color-singlet phase,
we can erase the $x$-dependent phase of the LOFF order parameter
and instead, the $x$-independent color-singlet term 
$\bar{\psi}\vec \gamma \!\cdot\! \vec q \psi$ is induced 
in the kinetic term for quarks.
Although the vector $\vec q$ is also gauge equivalent to 
the condensate $\VEV{\vec A_8}$, 
there is subtlety with respect to the tree gluon kinetic term:
Notice that the tree gluon potential does not give any contribution
to the free energy,
while it is relevant to the Meissner masses
for the transverse modes of the 4-7th gluons, 
because $T^8$ does not commute to $T^{4-7}$ and
thereby the corresponding Meissner masses are of the order of
$q (\equiv |\vec q|)$.
The point is that 
the values of $q$ are large 
in a wide range of the parameter region
where the LOFF phase exists, as we will see below.
In order to avoid confusion,
we may use the color-singlet transformation or
{\it we simply do not incorporate 
the tree gluon potential term into the LOFF phase in any case}.

\begin{figure}[t]
 \begin{center}
 \resizebox{0.47\textwidth}{!}{\includegraphics{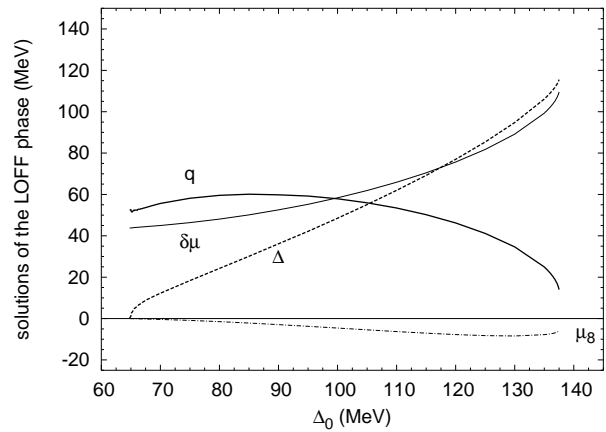}}
 \end{center}
 \caption{The dynamical solutions for the single plane wave LOFF phase.
          The bold solid, bold dashed, thin solid and
          thin dot-dashed curves represent the values of 
          $q \; (\equiv |\vec q|)$, $\Delta$, $\delta\mu$ and $\mu_8$, 
          respectively.
          The values $\Lambda=653.3$ MeV and $\mu=400$ MeV were used.
          \label{sol_loff}}
\end{figure}

\begin{figure}[t]
 \begin{center}
 \resizebox{0.47\textwidth}{!}{\includegraphics{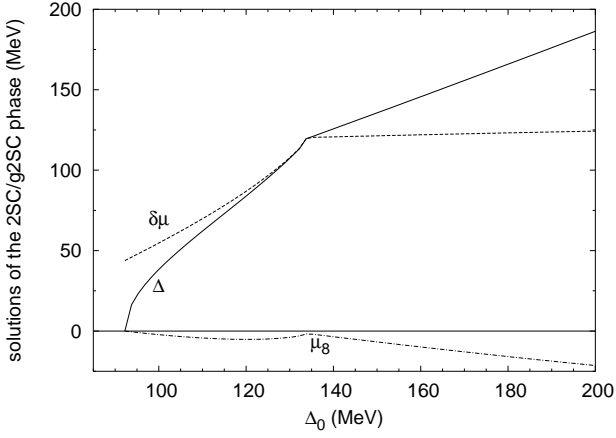}}
 \end{center}
 \caption{The dynamical solutions for the 2SC/g2SC phase.
          The solid, dashed and dot-dashed curves represent the values of 
          $\Delta$, $\delta\mu$ and $\mu_8$, respectively.
          At $\Delta_0=134.6$ MeV, 
          the g2SC phase turns into the 2SC one.
          The values $\Lambda=653.3$ MeV and $\mu=400$ MeV were used.
          \label{sol_2sc}}
\end{figure}

The dynamics of the minimal cylindrical gluonic phase~II is
analyzed in Ref.~\cite{Hashimoto:2007ut}
by solving the gap equations and the neutrality conditions 
in a self-consistent way.
We explicitly show the results for the gluonic, 
single plane wave LOFF and 2SC/g2SC phases
in Figs.\ref{sol_glu}--\ref{sol_2sc}, respectively.
In the analysis, we took realistic values $\mu=400$MeV and 
$\Lambda=653.3$MeV.
We also converted the four-diquark coupling constant $G_\Delta$
to the 2SC gap parameter $\Delta_0$ defined at $\delta\mu=0$
and varied the values of $\Delta_0$ from the weak coupling regime
($\Delta_0 \sim 60$ MeV) to the strong coupling one
($\Delta_0 \sim 200$ MeV).
For the gluonic phase, it is required to specify the value of 
$\alpha_s [\equiv g^2/(4\pi)]$,
although the results for the minimal cylindrical gluonic phase~II are not 
sensitive to the choice of the values of $\alpha_s$~\cite{Hashimoto:2007ut}.
We here took $\alpha_s=1$.

While the neutral normal phase without the diquark condensate 
always exists,
the neutral gluonic, LOFF, g2SC and 2SC phases do
only in the regions,
\begin{equation}
  \mbox{65.4MeV} < \Delta_0 < \mbox{160MeV}, \quad (\mbox{gluonic})
\end{equation}
\begin{equation}
  \mbox{64.9MeV} < \Delta_0 < \mbox{138MeV}, \quad (\mbox{LOFF})
\end{equation}
\begin{equation}
  \mbox{92.2MeV} < \Delta_0 < \mbox{134.6MeV}, \quad (\mbox{g2SC})
\end{equation}
and
\begin{equation}
 \Delta_0 >  \mbox{134.6MeV},  \quad (\mbox{2SC})
\end{equation}
respectively.

The analysis for the free energies has been done 
in Ref.~\cite{Hashimoto:2007ut}:
The normal phase is realized in the weak coupling regime with
$\Delta_0 < \mbox{64.9MeV}$.
While the single plane wave LOFF phase is energetically most favorable
only in the narrow region $\mbox{64.9MeV} < \Delta_0 < \mbox{67MeV}$,
the minimal cylindrical gluonic phase~II is stabler than 
the LOFF and 2SC/g2SC phases in the wide region 
$\mbox{67MeV} < \Delta_0 < \mbox{160MeV}$.
In the strong coupling regime with $\Delta_0 > \mbox{160MeV}$,
the 2SC phase is realized.

For the numerical calculation of the Meissner masses, 
we use the solutions shown in Figs.\ref{sol_glu}--\ref{sol_2sc}.
It is noticeable that the condensate $B$ 
is large 
in the almost whole region where the gluonic phase exists.
(See Fig.~\ref{sol_glu}.)
This feature is crucial for the Meissner masses in the gluonic phase, 
as we will see in the next section.

\section{Meissner masses in gluonic phase}
\label{formulae}
\subsection{Formulae}
\label{formulae-sub1}

Let us derive formulae for the numerical calculation of
the Meissner screening mass.

The squared Meissner mass can be expressed through
the second derivative of the effective potential:
\begin{eqnarray}
  \frac{\partial^2 V_{\rm eff}^R}{\partial A_\mu^a \partial A_\nu^b} &=&
  \Pi^{\mu\nu}_{\rm tree} 
  +\frac{g^2}{2}\int\frac{d^4 P}{i(2\pi)^4}
    \Tr \bigg[\,S_g \Gamma^{\mu a} S_g \Gamma^{\nu b}\,\bigg]
   \nonumber \\
&&-\frac{g^2}{2}\int\frac{d^4 P}{i(2\pi)^4}
    \Tr \bigg[\,S_g \Gamma^{\mu a} S_g \Gamma^{\nu b}\,\bigg]_{\rm c.t.},
  \label{d2V3}
\end{eqnarray}
where we defined the tree contribution
\begin{eqnarray}
  \Pi^{\mu\nu}_{\rm tree} &\equiv& \phantom{+}
    g^2 f^{a_1 a b}f^{a_1 a_2 a_3} A^{a_2\,\mu} A^{a_3\,\nu}
   \nonumber \\
&&+ g^2 f^{a_1 a a_2}f^{a_1 b a_3} g^{\mu\nu} A_\lambda^{a_2} A^{a_3\,\lambda}
   \nonumber \\
&&+ g^2 f^{a_1 a a_2}f^{a_1 a_3 b} A^{a_3\,\mu} A^{a_2\,\nu},
\end{eqnarray}
and the vertex
\begin{equation}
  \Gamma^{\mu a} \equiv g^{-1}\frac{\partial S_g^{-1}}{\partial A_\mu^a}
 =\left(\begin{array}{cc}
    \gamma^\mu T^a & 0 \\ 0 & -\gamma^\mu (T^a)^T
  \end{array}\right) \, . 
\end{equation}
In Eq.~(\ref{d2V3}), ``c.t.'' denotes the counter term and we used
\begin{equation}
  0=\frac{\partial }{\partial X}(S_g S_g^{-1}) = 
    \frac{\partial S_g}{\partial X} S_g^{-1} + 
    S_g \frac{\partial S_g^{-1}}{\partial X},
\end{equation}
for $X=A_\mu,\Delta,\delta\mu$,
and linearity of $S_g^{-1}$ with respect to all variables, i.e.,
$\frac{\partial^2 S_g^{-1}}{\partial X \partial Y}=0$.
We also abbreviated the bracket $\VEV{\cdots}$ for the gluon condensates.

For the numerical calculation, 
it is useful to diagonalize $S_g^{-1}$ and/or $S_g$.
Although the propagator inverse $S_g^{-1}$ in Eq.~(\ref{Sg-inv}) is a 
$48 \times 48$ matrix in the flavor, color, spinor and Nambu-Gor'kov
spaces,
we can block-diagonalize $S_g^{-1}$ in the flavor and chirality spaces.
This technique reduces our labour.

Let us transform the propagator inverse $S_g^{-1}$ as follows;
\begin{equation}
  S_g^{-1} (P)= 
  \left(
  \begin{array}{cc}
  \gamma^0 & 0 \\ 0 & i\varepsilon\gamma_5
  \end{array}
  \right)  
  \tilde{S}_g^{-1}(P)
  \left(
  \begin{array}{cc}
  1 & 0 \\ 0 & -i\varepsilon \gamma^0 \gamma_5 
  \end{array}
  \right), 
  \label{S_g}
\end{equation}
with
\begin{widetext}
\begin{equation}
  \tilde{S}_g^{-1}(P) = p_0 {\bf 1} + H_g, \qquad
  H_g \equiv 
  - \delta\mu\tau_3 +
  \left(
  \begin{array}{cc}
   \bar{\bar{\mu}} + \bm{\mu}_c
  - \gamma^0 \vec \gamma \cdot \vec p - \gamma^0 \vec \gamma \cdot \vec A
  & \epsilon^b\Delta \\ -\epsilon^b \Delta 
  &  - \bar{\bar{\mu}} - \bm{\mu}_c^T
  + \gamma^0 \vec \gamma \cdot \vec p - \gamma^0 \vec \gamma \cdot \vec A^T
  \end{array}
  \right) , 
\end{equation}
\end{widetext}
where
\begin{equation}
  \bar{\bar{\mu}} \equiv \mu -\frac{\delta\mu}{3}, \quad
  \bm{\mu}_c \equiv g A_0^a T^a, \quad
  \vec A \equiv g \vec A^a T^a \, .
\end{equation}
We here decomposed $\tilde{S}_g^{-1}$ into the diagonal $p_0$-part and 
the ``Hamiltonian'' $H_g$. 
(One can check easily hermiticity of $H_g$.)
Notice that the flavor dependence of $\tilde{S}_g^{-1}$ 
exists only in the first term of $H_g$ and therefore
$\tilde{S}_g^{-1}$ is flavor-diagonal.
Since the inverse of Eq.~(\ref{S_g}) yields the expression 
for the propagator,
\begin{equation}
  S_g (P) = 
  \left(
  \begin{array}{cc}
  1 & 0 \\ 0 & i\varepsilon \gamma^0 \gamma_5 
  \end{array}
  \right) 
  \tilde{S}_g (P)
  \left(
  \begin{array}{cc}
  \gamma^0 & 0 \\ 0 & i\varepsilon\gamma_5
  \end{array}
  \right)  ,
\end{equation}
the second derivative then reads
\begin{eqnarray}
  \frac{\partial^2 V_{\rm eff}^R}{\partial A_\mu^a \partial A_\nu^b} &=&
   \Pi^{\mu\nu}_{\rm tree}
  +\frac{g^2}{2}\int\frac{d^4 P}{i(2\pi)^4}
    \Tr \bigg[\,\tilde{S}_g \tilde{\Gamma}^{\mu a}
                \tilde{S}_g \tilde{\Gamma}^{\nu b}\,\bigg] 
   \nonumber \\
&&-\frac{g^2}{2}\int\frac{d^4 P}{i(2\pi)^4}
    \Tr \bigg[\,\tilde{S}_g \tilde{\Gamma}^{\mu a}
                \tilde{S}_g \tilde{\Gamma}^{\nu b}\,\bigg]_{\rm c.t.} ,
  \label{d2V3-2}
\end{eqnarray}
with 
\begin{eqnarray}
  \tilde{\Gamma}^{\mu a} &\equiv&
  \left(
  \begin{array}{cc}
  \gamma^0 & 0 \\ 0 & i\varepsilon\gamma_5
  \end{array}
  \right)  
  \Gamma^{\mu a}
  \left(
  \begin{array}{cc}
  1 & 0 \\ 0 & i\varepsilon \gamma^0 \gamma_5 
  \end{array}
  \right), \\[3mm]
&=&\left(\begin{array}{cc}
    \gamma^0 \gamma^\mu T^a & 0 \\ 0 & -\gamma^\mu \gamma^0 (T^a)^T
  \end{array}\right) \, . 
\end{eqnarray}
Since the current quark masses are ignored, 
the theory is chiral invariant.
Thus we can decompose the vertex and also the propagator 
into the right and left-handed parts.
One can easily show that the trace over the spinor space 
in Eq.~(\ref{d2V3-2}) is the sum of the two parts.

In virtue of hermiticity, we can diagonalize $H_g$ and $\tilde{S}_g$ 
by a unitary matrix $U$,
\begin{equation}
  H_g = U H_{\rm diag} U^\dagger, \quad 
  H_{\rm diag}=\diag(E_1^\tau,E_2^\tau,\cdots,E_n^\tau) , 
\end{equation}
\begin{equation}
  \tilde{S}_g = U
   \diag\left(\,\frac{1}{p_0+E_1^\tau},
                \frac{1}{p_0+E_2^\tau},\cdots,
                \frac{1}{p_0+E_n^\tau}\right) U^\dagger ,
  \label{S_g-diag}
\end{equation}
where $\tau=\pm$ for $\tau_3=\pm 1$ and
$E_{1,2,\cdots,n}^\tau$ denote the energy eigenvalues.
It is not difficult to find {\it numerically} the energy eigenvalues 
and the unitary matrix by using a standard method.
Noting that 
the integrand of Eq.~(\ref{d2V3-2}) contains the $p_0$-dependence
only in $\tilde{S}_g$ with the expression (\ref{S_g-diag}),
we can explicitly perform the integral over $p_0$ and thereby obtain
\begin{widetext}
\begin{eqnarray}
  \frac{\partial^2 V_{\rm eff}^R}{\partial A_\mu^a \partial A_\nu^b} &=&
   \Pi^{\mu\nu}_{\rm tree}
 -\frac{g^2}{2}\sum_{\tau=\pm}\sum_{E_i^\tau \ne E_j^\tau}\int\frac{d^3 p}{(2\pi)^3}
    \frac{\theta(E_i^\tau)-\theta(E_j^\tau)}{E_i^\tau-E_j^\tau}
      (U^\dagger \tilde{\Gamma}^{\mu a} U)_{ij}
      (U^\dagger \tilde{\Gamma}^{\nu b} U)_{ji} \nonumber \\
&&-\frac{g^2}{2}\sum_{\tau=\pm}\sum_{E_i^\tau=E_j^\tau}\int\frac{d^3 p}{(2\pi)^3}
      \delta(E_i^\tau)
      (U^\dagger \tilde{\Gamma}^{\mu a} U)_{ij}
      (U^\dagger \tilde{\Gamma}^{\nu b} U)_{ji} 
  \; -\mbox{(counter term)} \, .
  \label{d2V3-3}
\end{eqnarray}
\end{widetext}

We can derive similar formulae for other second derivatives, 
\begin{equation}
  \frac{\partial^2 V_{\rm eff}^R}{(\partial \Delta)^2}, \quad 
  \frac{\partial^2 V_{\rm eff}^R}{(\partial \mu_e)^2} , \quad
  \frac{\partial^2 V_{\rm eff}^R}{\partial \Delta \partial A_\mu^a}, \quad
   \mbox{etc..}
\end{equation}
It is also straightforward to extend the formulae to 
the version with a finite temperature. 

The formula (\ref{d2V3-3}) has an advantage over
the numerical derivative of the effective potential:
The numerical second derivative requires 
a quite precise calculation for the free energy and
thereby it takes a long time.
On the other hand, we can reach a sufficiently accurate result 
via (\ref{d2V3-3}) in a reasonable time.
Furthermore, in the expression of (\ref{d2V3-3}), 
it is clear that the contributions of the gapped and gapless modes 
to the Meissner masses are quite different.
(Compare the second and third terms in Eq.~(\ref{d2V3-3}).)
This might help us to understand why the existence of the gapless modes 
yields a sudden change of the squared Meissner mass for the 8th gluon 
at the border of the 2SC and g2SC phases~\cite{Huang:2004bg}.

\subsection{Numerical analysis}
\label{formulae-sub2}

\begin{figure}[tbp]
 \begin{center}
 \resizebox{0.47\textwidth}{!}{\includegraphics{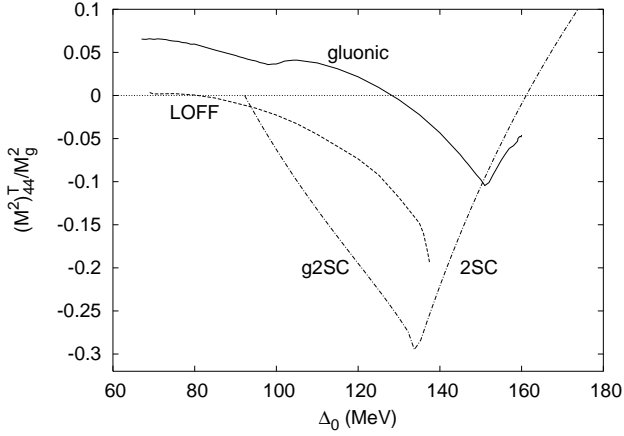}}
 \end{center}
 \caption{The squared Meissner masses of the transverse mode of 
          the 4th gluon in the unit of 
          $M^2_g\;[\equiv 4\alpha_s \mu^2/(3\pi)]$.
          The bold solid, dashed and dot-dashed curves are for
          the minimal cylindrical gluonic phase II, the single plane-wave LOFF
          and the 2SC/g2SC phases, respectively.
          The values $\Lambda=653.3$ MeV, $\mu=400$ MeV and
          $\alpha_s=1$ were used.
          \label{a4x}}
\end{figure}

Before the numerical calculation, we describe several features of 
the Meissner masses in the minimal cylindrical gluonic phase II.

In this phase,
the symmetry breaking structure is~\cite{Gorbar:2007vx}
\begin{eqnarray}
\lefteqn{\hspace*{-2cm}
 [SU(3)_c]_{\rm local} \times U(1)_{\rm em} \times SO(3)_{\rm rot} 
} \nonumber \\
&& \stackrel{\Delta,B}{\longrightarrow}
 \tilde{\tilde{U}}(1)_{\rm em} \times SO(2)_{\rm rot},
\end{eqnarray}
where the unbroken $\tilde{\tilde{U}}(1)_{\rm em}$ 
is connected with the new electric charge 
$\tilde{\tilde{Q}}_{\rm em}=Q_{\rm em}-\frac{1}{\sqrt{3}}T^8 - T^3$.
The rotational symmetry breaking leads to 
different Meissner masses for the transverse ($\mu,\nu=x,y$)
and longitudinal ($\mu,\nu=z$) modes. 
It is thus convenient to define
\begin{eqnarray}
  (M^2)_{ab}^T &\equiv& 
  \frac{\partial^2 V_{\rm eff}^R}{\partial A_x^a \partial A_x^b}=
  \frac{\partial^2 V_{\rm eff}^R}{\partial A_y^a \partial A_y^b},\\
  (M^2)_{ab}^L &\equiv& 
  \frac{\partial^2 V_{\rm eff}^R}{\partial A_z^a \partial A_z^b} \, .
\end{eqnarray}
Since we took the unitary gauge (\ref{U-gauge2}), 
the squared Meissner masses for the physical degrees of freedom are
\begin{equation}
  (M^2)_{11}^{T,L}=(M^2)_{22}^{T,L}, \quad  (M^2)_{44}^{T}=(M^2)_{55}^{T}, 
  \label{u1sym}
\end{equation}
and
\begin{equation}
  (M^2)_{33}^{T,L}, \quad (M^2)_{66}^{T,L}, \quad 
  (M^2)_{77}^{T}, \quad (M^2)_{88}^{T,L} ,
\end{equation}
where the relations (\ref{u1sym})
hold owing to the unbroken $\tilde{\tilde{U}}(1)_{\rm em}$ symmetry.
For the transverse modes of the 3rd and 8th gluons, 
it turns out that there exists a large mixing term $(M^2)_{38}^T$, so that 
we define the diagonal mass-squared terms for them,
\begin{equation}
  (M^2)_{33,{\rm diag}}^{T}, \quad (M^2)_{88,{\rm diag}}^{T} \, .
\end{equation}

Do there exist two Nambu-Goldstone (NG) bosons
connected with the symmetry breaking $SO(3)_{\rm rot} \to SO(2)_{\rm rot}$?
This is nontrivial because Goldstone's theorem for relativistically 
invariant theories is not necessarily valid
in noninvariant systems~\cite{Nielsen:1975hm,Miransky:2001tw}.
We find that the answer is formally ``yes'' in this case, 
as we will see below.

The point is that the rotational symmetry is spontaneously broken 
only by the condensate $\VEV{A_z^6} \ne 0$ 
in the minimal cylindrical gluonic phase II. 
Therefore, before taking the $z$-direction, the effective potential 
should depend on the $SO(3)_{\rm rot}$ invariant
\begin{equation}
  {\cal B} \equiv \sum_{i=x,y,z} \VEV{A_i^6}^2 , 
\end{equation}
i.e.,
\begin{equation}
  V_{\rm eff}^R = V_{\rm eff}^R({\cal B}) \, .
\end{equation}
We then find 
\begin{equation}
 \frac{\partial^2 V_{\rm eff}^R}{\partial \VEV{A_i^6} \partial \VEV{A_j^6}}=
   2 \delta_{ij}\frac{\partial V_{\rm eff}^R}{\partial {\cal B}}
 + 4 \VEV{A_i^6} \VEV{A_j^6}
     \frac{\partial^2 V_{\rm eff}^R}{(\partial {\cal B})^2} \, .
\end{equation}
Since we take the direction $\VEV{A_{x,y}^6}=0$, $\VEV{A_z^6} \ne 0$
and the gap equation for $\VEV{A_z^6}$ yields 
$\frac{\partial V_{\rm eff}^R}{\partial {\cal B}}=0$, 
we formally find that the squared Meissner mass for the transverse mode of 
the 6th gluon is vanishing,
\begin{equation}
  (M^2)_{66}^T = 0 \, . 
  \label{NG6}
\end{equation}
It implies that $A_{x,y}^6$ correspond to the two NG bosons.

\begin{figure}[t]
 \begin{center}
 \resizebox{0.47\textwidth}{!}{\includegraphics{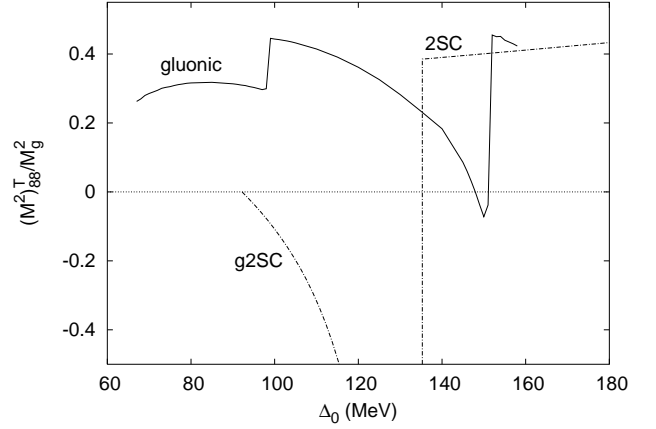}} 
 \end{center}
 \caption{The squared Meissner masses of the transverse mode of 
          the 8th gluon in the unit of 
          $M^2_g\;[\equiv 4\alpha_s \mu^2/(3\pi)]$.
          The bold solid and dot-dashed curves are for
          the minimal cylindrical gluonic phase II and the 2SC/g2SC phases, 
          respectively.
          The values $\Lambda=653.3$MeV, $\mu=400$MeV and  
          $\alpha_s=1$ were used.
          In the g2SC phase, the squared Meissner mass monotonously 
          decreases and goes to minus infinity at $\Delta_0=134.6$MeV, 
          although it is not explicitly shown here.
          \label{a8x}}
\end{figure}

A crucial difference between the gluonic and 2SC/g2SC phases
is the existence of the tree gluon potential term.
Although the effect is negligible for the free energy 
in the minimal cylindrical gluonic phase II,
it is quite important for the Meissner masses.
Neglecting the suppressed terms 
$\sim {\cal O}(\mu_3^2),{\cal O}(\mu_3 \mu_8),{\cal O}(\mu_8^2)$,
we obtain the tree terms of the squared Meissner masses:
\begin{subequations}
\label{M2T}
\begin{align}
&(M^2)_{11,22}^T \simeq (M^2)_{44,55}^T \simeq (M^2)_{33}^T = \frac{B^2}{4}, \\
&(M^2)_{77}^T \simeq B^2, \qquad (M^2)_{88}^T = \frac{3}{4}B^2, \\
&(M^2)_{38}^T = -\frac{\sqrt{3}}{4}B^2 \, . 
\end{align}
\end{subequations}
Thus the transverse modes except for 
$(M^2)_{66}^T$ and $(M^2)_{33,{\rm diag}}^{T}$
have the {\it positive and large} contributions of the order of $B^2$.
(For the values of $B$, see Fig.~\ref{sol_glu}.)
This is one of the reasons why the Meissner masses tend to be
positive compared with those in the 2SC/g2SC phase.

\begin{figure}[tbp]
 \begin{center}
 \resizebox{0.47\textwidth}{!}{\includegraphics{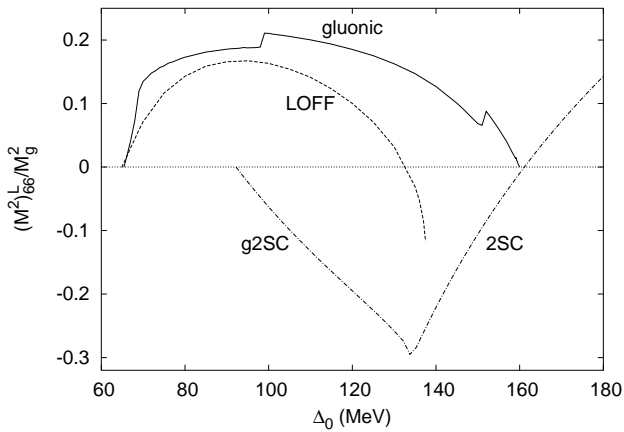}}
 \end{center}
 \caption{The squared Meissner masses of the longitudinal mode of 
          the 6th gluon in the unit of 
          $M^2_g\;[\equiv 4\alpha_s \mu^2/(3\pi)]$.
          The bold solid, dashed and dot-dashed curves are for
          the minimal cylindrical gluonic phase II, the single plane-wave LOFF
          and the 2SC/g2SC phases, respectively.
          The values $\Lambda=653.3$MeV, $\mu=400$MeV and 
          $\alpha_s=1$ were used.
          \label{a6z}}
\end{figure}

We also note some features of the Meissner masses for 
the single-plane wave 2SC-LOFF phase. 
The $A_\mu^{1-3}$ gluons should be massless
and the relations
\begin{equation}
  (M^2)_{44}^{T,L}=(M^2)_{55}^{T,L}=(M^2)_{66}^{T,L}=(M^2)_{77}^{T,L}
\end{equation}
hold because of the unbroken $SU(2)_c$ gauge symmetry.
In addition, similarly to (\ref{NG6}), we formally obtain
\begin{equation}
  (M^2)_{88}^T = 0 \, .
  \label{NG8}
\end{equation}

Let us now turn to the numerical analysis of the Meissner masses.

\begin{figure}[tbp]
 \begin{center}
 \resizebox{0.47\textwidth}{!}{\includegraphics{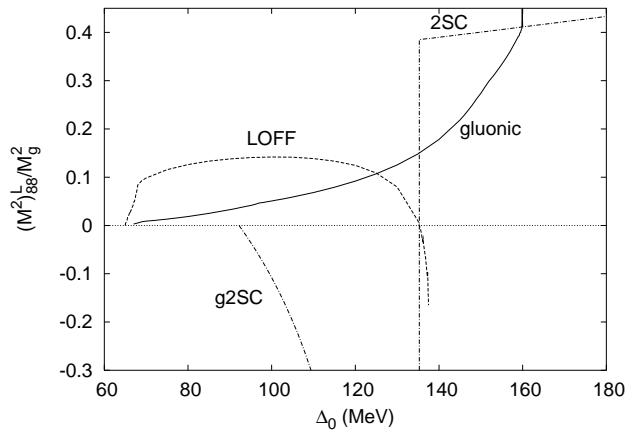}} 
 \end{center}
 \caption{The squared Meissner masses of the longitudinal mode of 
          the 8th gluon in the unit of 
          $M^2_g\;[\equiv 4\alpha_s \mu^2/(3\pi)]$.
          The bold solid, dashed and dot-dashed curves are for
          the minimal cylindrical gluonic phase II, the single plane-wave LOFF
          and the 2SC/g2SC phases, respectively.
          The values $\Lambda=653.3$MeV, $\mu=400$MeV and
          $\alpha_s=1$ were used.
          As in Fig.\ref{a8x}, the curve for the g2SC phase 
          below the figure is cut off.
          \label{a8z}}
\end{figure}

\begin{figure}[tbp]
 \begin{center}
 \resizebox{0.47\textwidth}{!}{\includegraphics{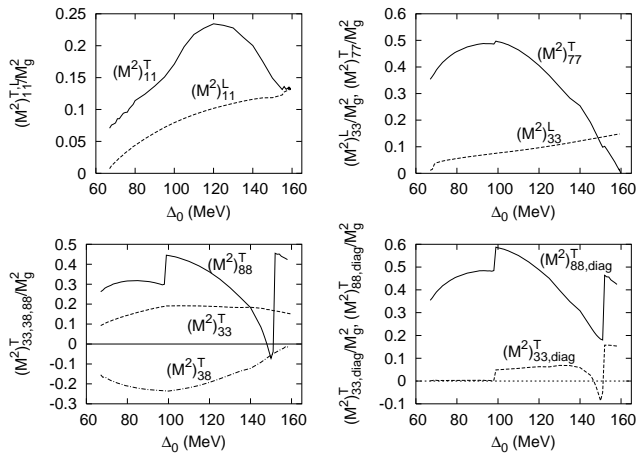}} 
 \end{center}
 \caption{The squared Meissner masses for several gluonic modes
          in the minimal cylindrical gluonic phase II
          in the unit of 
          $M^2_g\;[\equiv 4\alpha_s \mu^2/(3\pi)]$.
          The values $\Lambda=653.3$MeV, $\mu=400$MeV and
          $\alpha_s=1$ were used.
          In the right bottom figure, we showed the diagonalized mass
          squared terms for the transverse modes of the 3rd and 8th gluons.
          \label{a1-8}}
\end{figure}

We depict the results in Figs.\ref{a4x}--\ref{a1-8} in the unit of
$M_g^2 [\equiv 4\alpha_s \mu^2/(3\pi)]$.
In the analysis, we used $\mu=400$MeV, $\Lambda=653.3$MeV
and $\alpha_s=1$.

For the minimal cylindrical gluonic phase II,
the squared Meissner masses $(M^2)_{44}^T=(M^2)_{55}^T$
are positive in the region $\mbox{65.4MeV} < \Delta_0 < \mbox{130MeV}$,
while it suffers from the chromomagnetic instability in 
$\mbox{130MeV} < \Delta_0 < \mbox{160MeV}$. (See Fig.\ref{a4x}.)
We also find that $(M^2)_{88}^T$ becomes negative 
in the small region around $\Delta_0 \sim 150$MeV.
(See Fig.\ref{a8x}.)
For the other modes, however, the chromomagnetic instability
does not occur as shown in Figs.\ref{a6z}--\ref{a1-8}.
(After the diagonalization of $(M^2)^{T}_{33}$, $(M^2)^{T}_{38}$ 
and $(M^2)^{T}_{88}$, the instability in $(M^2)^{T}_{88}$ is converted into 
$(M^2)^{T}_{33,{\rm diag}}$, because we define the diagonalized
squared masses as $(M^2)^{T}_{33,{\rm diag}} < (M^2)^{T}_{88,{\rm diag}}$.)
It is quite noticeable that there exist spikes and valleys 
around $\Delta_0 \sim 100$MeV and
$\Delta_0 \sim 150$MeV in Figs.\ref{a4x}--\ref{a6z}.

How about the sensitivity of the Meissner masses on $\alpha_s$?
Although the dynamical solutions of $\Delta$, $B$, $\delta\mu$, $\mu_3$
and $\mu_8$ in the minimal cylindrical gluonic phase II
are almost independent of $\alpha_s$~\cite{Hashimoto:2007ut},
the Meissner masses for the transverse modes can be sensitive. 
Note that the one-loop contributions are proportional to $\alpha_s$
and thus the influence of the tree contributions (\ref{M2T}) is 
relatively stronger (weaker) as the values of $\alpha_s$ decrease (increase).
For example, $(M^2)_{44}^T$ becomes negative at
$\Delta_0=140,130,120$MeV for $\alpha_s=0.85,1.0,1.15$, respectively.
For the other transverse modes in Eq.~(\ref{M2T}), 
there should appear similar sensitivities.
On the other hand, for the longitudinal modes, 
the ratio $(M^2)_{ab}^L/M_g^2$ is insensitive to $\alpha_s$, 
because the tree contributions are suppressed.

For the single plane wave LOFF phase, 
like in the minimal cylindrical gluonic phase II,
the transverse modes of the 4-7th gluons are the most problematic.
However the chromomagnetic instability occurs in the earlier region,
$\mbox{80MeV} < \Delta_0 < \mbox{138MeV}$.
(See Fig.\ref{a4x}.)
All the other modes also suffer from the chromomagnetic instability
in the end of the LOFF phase around $\Delta_0 \sim 130$MeV. 
(See Fig.\ref{a6z} and \ref{a8z}.)
This is different from the situation 
in the minimal cylindrical gluonic phase II.
We also note that these results are consistent with 
the analysis based on the HDL 
approximation~\cite{Giannakis:2004pf,Gorbar:2005tx}.

The 2SC/g2SC phase has the chromomagnetic instability 
numerically in the region $\mbox{92.2MeV} < \Delta_0 < \mbox{160MeV}$.
The results agree with those in Appendix B in the second paper of
Ref.~\cite{Huang:2004bg} (, see also Refs.~\cite{Kiriyama:2006xw,He:2006vr}).

In conclusion, although the minimal cylindrical gluonic phase II
does not completely remove the chromomagnetic instability,
it cures the situation in a wide region.

The comments about the massless modes are in order.
The numerical calculations including the non-HDL effects 
do not necessarily reproduce the vanishing Meissner masses.
This fact is known even in the 2SC phase, i.e., the $A_\mu^{1-3}$ gluons
acquire the non-HDL contributions like $\Delta^2 \log \Lambda^2/\Delta^2$ 
in the sharp-cutoff regularization scheme~\cite{Rischke:2000qz,Alford:2005qw}.
In order to settle this problem, a more sophisticated regularization 
scheme is required. 
It will be studied elsewhere.

\section{Summary and discussions}
\label{summary}

We analyzed the Meissner screening masses in the simplest gluonic phase
(the minimal cylindrical gluonic phase II)
as well as the single plane wave LOFF and 2SC/g2SC phases.
We derived the formulae for the Meissner masses without 
any help of the numerical derivative.
It was found that in the formulae the gapless mode makes the contribution
characterized by the Dirac's $\delta$-function.
We showed that the simplest gluonic phase removes 
the chromomagnetic instability in the region
$\mbox{65.4MeV} < \Delta_0 < \mbox{130MeV}$,
whereas the single plane wave LOFF one does in
$\mbox{64.9MeV} < \Delta_0 < \mbox{80MeV}$.
We here took
$\Lambda=653.3$MeV, $\mu=400$MeV and $\alpha_s=1$.
The 2SC phase does not have the chromomagnetic instability 
in the strong coupling regime $\Delta_0 > \mbox{160MeV}$.
Incorporating the analysis of the free energy~\cite{Hashimoto:2007ut},
we conclude that in the region $\mbox{67MeV} < \Delta_0 < \mbox{130MeV}$
the simplest ansatz for the gluonic phases works more nicely
than the single plane wave LOFF and g2SC phases.
On the other hand, 
the single plane wave LOFF phase is energetically more favorable 
and also resolves the chromomagnetic instability
only in the window $\mbox{64.9MeV} < \Delta_0 < \mbox{67MeV}$.

Furthermore, we found the noticeable behaviours of 
the squared Meissner masses in the minimal cylindrical gluonic phase II
around $\Delta_0 \sim 100$MeV and $\Delta_0 \sim 150$MeV.
(See Figs.\ref{a4x}--\ref{a6z}.)
The 2SC/g2SC phase also has the similar behaviours:
Notice that there appears the abrupt change of the Meissner mass
for the 8th gluon from the g2SC side to the 2SC one
and that the values of the squared Meissner mass for the 4-7th gluons
have the valley at the phase transition point from the g2SC phase to 
the 2SC one, as shown in Figs.\ref{a4x}--\ref{a8z}.
These similarities might suggest that new gapless modes 
in the minimal cylindrical gluonic phase II appear
around $\Delta_0 \sim 100$MeV and $\Delta_0 \sim 150$MeV.
We also note that the value of $B$ takes its maximum 
around $\Delta_0 \sim 100$MeV and that the relation 
$B \simeq \delta\mu$ is satisfied around $\Delta_0 \sim 150$MeV.
(See Fig.\ref{sol_glu}.)
These facts might be significant.
The dispersion relation for quarks in the gluonic phase
will be performed elsewhere.

There still exists a chromomagnetically unstable region. 
However it should be noticed that we examined only the simplest 
ans\"{a}tze for the LOFF and gluonic phases in this paper.
The multiple plane wave LOFF phase may completely remove 
the chromomagnetic instability in 
the whole parameter region~\cite{Bowers:2002xr,Gatto:2007ja}.
More involved gluonic phases can also resolve the instability:
Since there appears the illness in the transverse mode 
of the 4th gluon, the GCSL phase with the gluon condensates
$\mu_8=\sqrt{3}/2 g\VEV{A_0^8}$ and
$K=g\VEV{A_y^4}=g\VEV{A_z^6}$~\cite{Gorbar:2007vx,Hashimoto:2007ut} 
is hopeful, for example.
An important point is that the free energy for the GCSL phase is 
slightly lower than that for the minimal cylindrical gluonic 
phase~II~\cite{Hashimoto:2007ut}.
Inhomogeneous gluonic phases are also 
interesting~\cite{Gorbar:2006up,Ferrer:2007uw}.

Independently of the chromomagnetic instability, 
the Sarma instability for the diquark Higgs mode
should be removed as well.
This problem will be considered elsewhere.

\acknowledgments
 
The authors thank V.~A.~Miransky for fruitful discussions.
J.J. acknowledges useful discussions with Razvan Nistor.
The numerical calculations were carried out on Altix3700 BX2 
at YITP in Kyoto University.
The work was supported by the Natural Sciences and Engineering
Research Council of Canada.

\end{document}